\newif\iffinal
\finalfalse	
\finaltrue 

\newif\ifmarek
\marektrue

\documentclass[reqno,twoside,11pt]{amsart}
\iffinal\else\usepackage[notref,notcite]{showkeys}\fi
\usepackage{cite}
\usepackage{subcaption}
\usepackage{amsmath}
\usepackage{amsfonts}
\usepackage{amssymb}
\usepackage{graphicx}
\usepackage{enumitem}
\usepackage{graphics}
\usepackage{tikz}
\usetikzlibrary{arrows,automata,positioning}
\graphicspath{{/Users/aubain2015/Documents/york/}}
\usepackage{verbatim}
\usepackage{color}
\usepackage{float}
\usepackage{dirtytalk}
\IfFileExists{epsf.def}{\input epsf.def}{\usepackage{epsf}}

\ifmarek
\IfFileExists{myowntimes.sty}{\usepackage{myowntimes}\usepackage{mathrsfs}}
	{\usepackage{times}\newcommand{\mathscr}{\mathcal}}

\usepackage{mathptmx}
\fi

\DeclareFontFamily{OT1}{eusb}{} \DeclareFontShape{OT1}{eusb}{m}{n}
{<5> <6> <7> <8> <9> <10> <11> <12> <14.4> eusb10}{}
\DeclareMathAlphabet{\eusb}{OT1}{eusb}{m}{n}

\DeclareFontFamily{OT1}{eusm}{} \DeclareFontShape{OT1}{eusm}{m}{n}
{<5> <6> <7> <8> <9> <10> <11> <12> <14.4> eusm10}{}
\DeclareMathAlphabet{\eusm}{OT1}{eusm}{m}{n}

\DeclareFontFamily{OT1}{eufm}{} \DeclareFontShape{OT1}{eufm}{m}{n}
{<5> <6> <7> <8> <9> <10> <11> <12> <14.4> eufm10}{}
\DeclareMathAlphabet{\mathfrak}{OT1}{eufm}{m}{n}

\DeclareFontFamily{OT1}{fraktura}{}
\DeclareFontShape{OT1}{fraktura}{m}{n} {<5> <6> <7> <8> <9> <10>
<11> <12> <13> <14.4> [1.1] eufm10}{}
\DeclareMathAlphabet{\fraktura}{OT1}{fraktura}{m}{n}

\DeclareFontFamily{OT1}{cmfi}{} \DeclareFontShape{OT1}{cmfi}{m}{n}
{<5> <6> <7> <8> <9> <10> <11> <12> <13> <14.4> [0.9] cmfi10}{}
\DeclareMathAlphabet{\cmfi}{OT1}{cmfi}{b}{n}

\DeclareFontFamily{OT1}{cmss}{} \DeclareFontShape{OT1}{cmss}{m}{n}
{<5> <6> <7> <8> <9> <10> <11> <12> <13> <14.4> cmss10}{}
\DeclareMathAlphabet{\cmss}{OT1}{cmss}{m}{n}

\newtheoremstyle{thm}{1.5ex}{1.5ex}{\itshape\rmfamily}{}
{\bfseries\rmfamily}{}{2ex}{}

\newtheoremstyle{def}{1.5ex}{1.5ex}{\slshape\rmfamily}{}
{\bfseries\rmfamily}{}{2ex}{}

\newtheoremstyle{rem}{1.3ex}{1.3ex}{\rmfamily}{}
{\itshape}
{} {1.5ex}{}

\theoremstyle{thm}
\newtheorem{theorem}{Theorem}[section]
\newtheorem{lemma}[theorem]{Lemma}

\newtheorem{corollary}[theorem]{Corollary}

\theoremstyle{def}

\theoremstyle{rem}

\numberwithin{equation}{section}

\renewcommand{\section}{\secdef\sct\sect}
\newcommand{\sct}[2][default]{\refstepcounter{section}
\addcontentsline{toc}{section}
{{\tocsection {}{\thesection}{\!\!\!\!#1\dotfill}}{}}
\vspace{0.7cm}
\centerline{ 
\scshape\arabic{section}.\ #1} \nopagebreak \vspace{0.2cm}}
\newcommand{\sect}[1]{
\vspace{0.4cm} \centerline{\large\scshape\rmfamily #1}
\vspace{0.2cm}}

\renewcommand{\subsection}{\secdef\subsct\sbsect}
\newcommand{\subsct}[2][default]{\refstepcounter{subsection}
\addcontentsline{toc}{subsection}
{{\tocsection{\!\!}{\hspace{1.2em}\thesubsection}{\!\!\!\!#1\dotfill}}{}}
\nopagebreak\vspace{0.45\baselineskip} {\flushleft\bf
\thesection.\arabic{subsection}~\bf #1.~}
\\*[3mm]\noindent
\nopagebreak}
\newcommand{\sbsect}[1]{\vspace{0.1cm}\noindent
\textbf{#1.~}\vspace{0.1cm}}

\renewcommand{\subsubsection}{%
\secdef \subsubsect\sbsbsect}
\newcommand{\subsubsect}[2][default]{%
\refstepcounter{subsubsection} 
\addcontentsline{toc}{subsubsection}{{\tocsection{\!\!}
{\hspace{3.05em}\thesubsubsection}{\!\!\!\!#1\dotfill}}{}}
\nopagebreak
\vspace{0.15\baselineskip} \nopagebreak {\flushleft\rmfamily
\itshape\arabic{section}.\arabic{subsection}.\arabic{subsubsection}
\ \rmfamily #1\/.}\ }
\newcommand{\sbsbsect}[1]{\vspace{0.1cm}\noindent
\rmfamily \itshape
\arabic{section}.\arabic{subsection}.\arabic{subsubsection} \
\sffamily #1\/.\ }

\iffinal

\else

\fi

\newcommand{\N}{\mathbb N}

\newcommand{\R}{\mathbb R}

\title[]
{\large SIS Epidemic Model\\ Birth-and-Death Markov Chain Approach}

\author[]{Aubain Nzokem$^1$}
\thanks{$^1$ aubain14@yorku.ca}
\begin{document}

\maketitle

\vspace{-2mm}
\centerline{\textit{Department of Mathematics \& Statistics, York University, Toronto}}

\vspace{-2mm}
\begin{abstract}
We are interested in describing the infected size of the SIS Epidemic model using Birth-Death Markov process. The Susceptible-Infected-Susceptible (SIS) model is defined within a population of constant size $M$; the size is kept constant by replacing each death with a newborn healthy individual. The life span of each individual in the population is modelled by an exponential distribution with parameter $\alpha$; and the disease spreads within the population is modelled by a Poisson process with a rate $\lambda_{I}$. $\lambda_{I}=\beta I(1-\frac{I}{M}) $ is similar to the instantaneous rate in the logistic population growth model. The analysis is focused on the disease outbreak, where the reproduction number $R=\frac{\beta} {\alpha} $ is greater than one. As methodology, we use both numerical and analytical approaches. The analysis relies on the stationary distribution for Birth and Death Markov process. The numerical approach creates sample path simulations into order show the infected size dynamics, and the relationship between infected size and $R$. As $M$ becomes large, some stable statistical characteristics of the infected size distribution can be deduced. And the infected size is shown analytically to follow a normal distribution with mean $(1-\frac{1}{R}) M$ and Variance $\frac{M}{R} $.\\

\end{abstract}

 \section{Introduction}
\noindent
The Birth and Death Markov Chain is a special class of the continuous stochastic process. The importance of such class arises from the fact that it is generated by combining two standard processes (Birth process and Death process). The stationary distribution of such process at the equilibrium was studied in the Mathematics literature\cite {karlin1975}. The findings are useful as one of the interests in studying stochastic process is to describe the behaviour of the stochastic process in the long run; that is, how the process is distributed when the time becomes large.\\
The Susceptible-Infected-Susceptible (SIS) model is one of the simplest and most paradigmatic models in mathematical epidemiology. The stochastic version of the SIS model was studied by N{\aa}sell\cite {naasell1999quasi,naasell1996quasi}, who is among the  pioneers  to report the normal distribution nature of the quasi-stationary distribution when the reproduction number is greater than one and  the population size ($M$) is large. The major critic was his methodology. Ovaskainen\cite {ovaskainen2001} argues that the methodology was heuristic. In general, the literature review\cite {aubain2020, naasell1996quasi, allen2008introduction, ovaskainen2001, clancy2011, wierman2004} offers many approximations of the quasi-stationary distribution of the SIS model, which may reflect the difference in methodologies or in parametrisations. Some studies\cite {wierman2004} consider transmission parameter ($\beta$) as a function of the population size ($M$).\\
In our study, the Birth and Death Markov chain is used to describe the dynamics of the SIS model and its features; the parameter are fixed, and the reproduction number ($R$) is greater than one, only the size of the population $(M)$ can change. In the next section, we will formulate the disease spreading parameters and analyze some infected size sample paths and how they are impacted the reproduction number. And the last section will focus on the distribution of the infected size. Both numerical and analytical approaches will be used to analyse the distribution nature of the asymptotic infected size.

\section{Epidemic Spreading: Death and Infection Process Modelling}

\noindent
In SIS model, the population is divided at each time $t$ into susceptible individuals ($S(t)$) and infective individuals ($I(t)$). The evolution of these quantities is usually described in Epidemiology by the following deterministic differential equations (\ref{eq1}) $(a)$:
\begin{flalign}
  \begin{aligned}
& (a)            &     & (b)\\
\frac{dS}{dt}&=-\frac{\beta} {M}SI+\alpha I   &\quad \frac{S}{M} &=\frac{\alpha} {\beta} =\frac{1}{R} \\
\frac{dI}{dt}&=\frac{\beta} {M}SI-\alpha I    & \frac{I}{M} &=1-\frac{\alpha} {\beta} =1-\frac{1}{R}
\label{eq1}
  \end{aligned}
\end{flalign}
\noindent
 The parameters $\beta$ and $\alpha$ are respectively the transmission rate and the rate of death and birth. To have population size ($M=S+I$) constant over time, each individual who dies is replaced by a susceptible individual. The threshold value $R=\frac{\beta} {\alpha} $, which is a basic Reproduction Number, is an indicator that determines whether we will have extinction of the disease ($0<R<1$) or an outbreak of the disease ($R>1$). One of the results of the deterministic differential equations is the equilibrium (\ref{eq1}) $(b)$ of the system (\ref{eq1}) $(a)$ in the long run when $R>1$. At the equilibrium (\ref{eq1}) $(b)$, the portion infected is constant ($1-\frac{1}{R} $).\\
\noindent
The deterministic version of the SIS model was introduced by Kermack and McKendrick and has been fully analysed. For related deterministic work of interest, see \cite {aubain2020, Nzokem2020EpidemicDA, aubain2021}.\\
The stochastic version, called the stochastic logistic
epidemic model, is usually modelled as Birth and Death Markov Chain where the transition probability is defined as follows for I(t) taking value on $\{0,1, \dots, M\} $ 
\begin{equation}
 P_ {I, J} (t, t+h) =
  \begin{cases}
    \lambda_{I}h + o_1(h) & \quad \text {if} J=I+1\\
    \mu_{I}h + o_2(h) & \quad \text {if} J=I-1\\
 1-(\lambda_{I}+\mu_{I}) h +o_3(h) & \quad \text {if} J=I
  \end{cases}
   \label {eq: l11}
\end{equation}

with $\lim_ {h\rightarrow 0} o_1(h)=0; \lim_ {h\rightarrow 0} o_2(h)=0; \lim_ {h\rightarrow 0} o_3(h)=0$\\
\noindent
We suppose the transmission rate $\beta=c\theta$ can be written as a product of contact rate $(c)$ and the probability of infection $(\theta) $. The transition probability $(\ref {eq: l11}) $ can be fully determined by the following set of rules\cite{clancy2011}: (a) each individual gets into contact with another individual after an elapsed time, which follows an independent and identically distribution (iid). The elapsed time is exponentially distributed with parameter $c$. And if the contact involves a susceptible individual and infected individual, the probability of infection is $\theta$. The transmission rate is $\beta=c\theta$. (b) the infected lifetime is also an exponentially distributed with parameter $\alpha$, because of the memoryless property of the lifetime.\\

\noindent
For $J=I+1$ in $(\ref {eq: l11}) $, all I infected individuals get into contact with another individual according to a Poisson process with parameter $cI$, since $(1-\frac{I}{M}) $ is the probability to meet susceptible individuals and $\theta$ is the probability of infection, by thinning the Poisson
process, we conclude that the contacts between infected and susceptible individuals that will end up with infection follows a Poisson process with parameter $\lambda_{I}=\beta I(1-\frac{I}{M}) $.\\
 For $(J=I-1) $ in $(\ref {eq: l11}) $, the number of infected individuals that becomes susceptible individuals follows a Poisson process with parameter $\mu_I=\alpha I$.
 \subsection {Infected Size Samples}
\begin{figure}[ht]
     \centering
         \centering
         \includegraphics[scale=0.6] {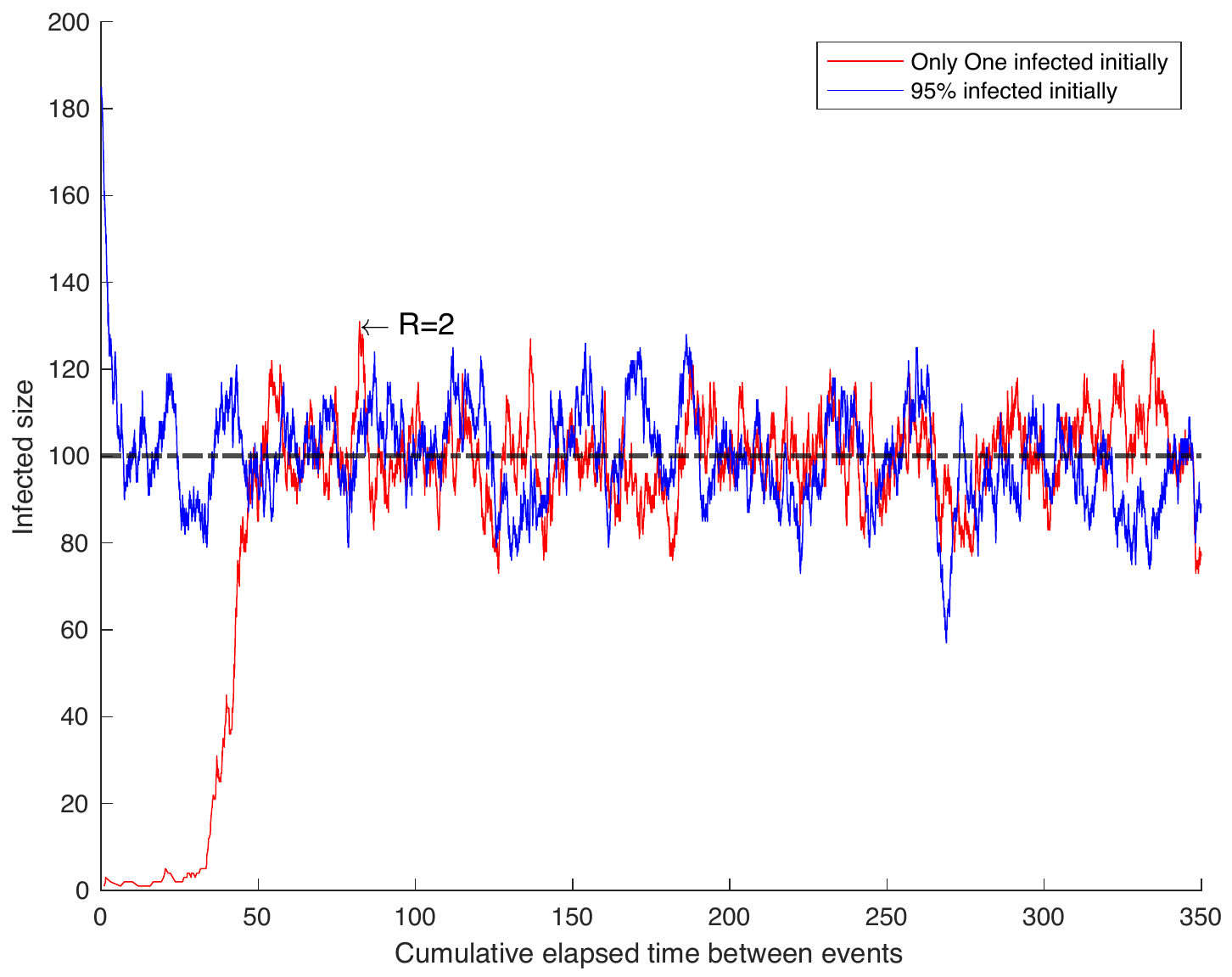}
        \caption {Deterministic versus Stochastic equilibrium of the infected size}
        \label{fig1}
\end{figure}

\noindent 
Based on the assumptions and parameters developed previously, a MATLAB program with $M$ individuals was created, and the main variables were age, health status, cumulative elapsed of time between events. At the death of an individual, a healthy individual and his life span are introduced in the program code. By controlling the age, we can focus on the infection process over the time.\\
For $R=2$ and $\alpha=0.3$, we have the infected size from two variables:  heathy status and cumulative elapsed of time between infections. Two sample paths are presented in Fig \ref{fig1} with only one infected initially and with $95\%$ infected initially.

\noindent
In the case of only one infected initially, the infected size grows at an increasing rate before fluctuating around the deterministic equilibrium $M*(1-\frac{1}{R}) $ developed in (\ref{eq1}) $(b)$. In the second case of $95\%$ infected initially, as shown in Fig \ref{fig1}, the infected size decreases rapidly before fluctuating around the deterministic equilibrium.\\

\begin{figure}[ht]
     \centering
         \centering
         \includegraphics[scale=0.6] {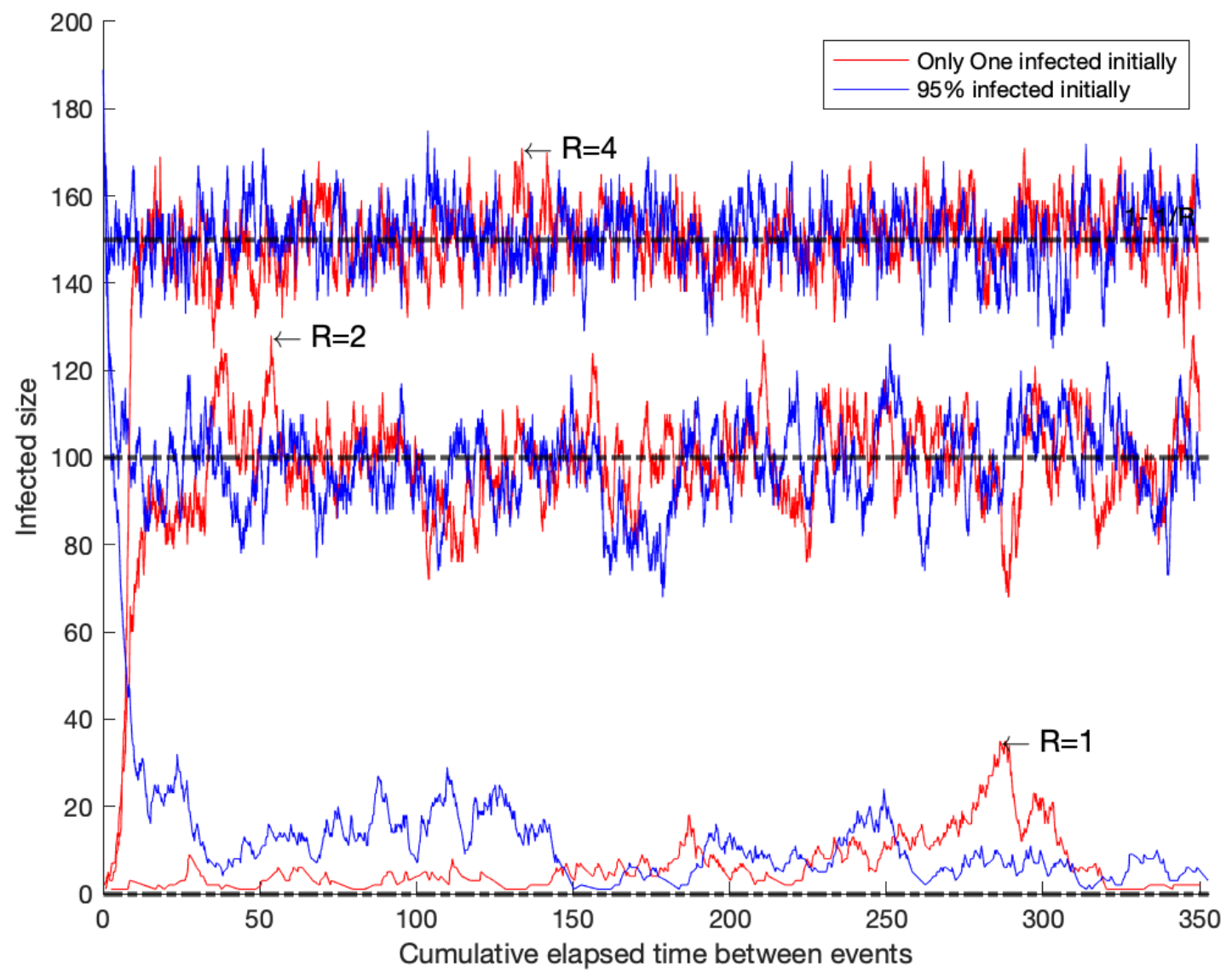}
        \caption {Impact of the Reproduction Number ($R$) on the equilibrium of the stochastic infected size}
        \label{fig2}
\end{figure}

\noindent
We have illustrated in Fig \ref{fig2} how the sample path reacts with respect to the reproduction number (R). In fact, when there is only one infected initially and the reproduction number is greater than 1, the infected size increases rapidly, before fluctuating around deterministic equilibrium $M*(1-\frac{1}{R})$, straight line in black color in Fig \ref{fig2}. The same pattern is observed when there is $95\%$ infected initially, the fluctuation follows a rapidly decreasing. The stability is also shown in Fig \ref{fig2} for $R=2$ and $R=4$, whereas for $R=1$, the process is unstable, and the disease will eventually die out.

\noindent
In addition, for $R\geq1$, there is a positive relation between the infected size and the reproduction number ($R$). As illustrated in Fig $\ref{fig2} $, the infected size increases when $R$ increases. 

\subsection{Infected Size Stationary Distribution}
\noindent
Based on the parameters developed in $(\ref {eq: l11}) $, we have a Birth and Death process on the state space $\{0,1,\ldots,M\}$ with transition rates:
\begin{align*}
 \lambda_k = \beta\left(\frac{M-k} {M}\right) k  \hspace{5mm}(k\rightarrow k+1)  \quad \quad
  \mu_k=\alpha k \hspace{5mm} (k\rightarrow k-1)
\end{align*}
\begin{tikzpicture}[shorten >=1pt,node distance=2cm,on grid,auto,scale=0.5]
  \tikzstyle{every state}=[fill={rgb:black,2;white,10}]
    \node[state, initial]   (q_1)                   {$0$};
    \node[state]           (q_2) [right of=q_1]    {$1$};
    \node[state]           (q_3) [right of=q_2]    {$2$};
    \node[state]           (q_4) [right of=q_3]    {$3$};
    \node                  (q_d) [right of=q_4]    {$\cdots\cdots$}; 
    \node[state]           (q_5) [right of=q_d]    {$M-1$};    
    \node[state]           (q_6) [right of=q_5]    {$M$};
        
    \path[->]
    (q_1) edge [loop above] node { }    			   (   )
       
    (q_2) edge [bend left]  node {$\mu_{1}$}       (q_1)
          edge [bend left]  node {$\lambda_{1}$}   (q_3)
          edge [loop above] node { }    			   (   )
    (q_3) edge [bend left]  node {$\mu_{2}$}       (q_2)
          edge [bend left]  node {$\lambda_{2}$}   (q_4)
          edge [loop above] node { }               (   )
    (q_4) edge [bend left]  node {$\mu_{3}$}       (q_3)
          edge [loop above] node { }               (   )
          edge [bend left]  node {$\lambda_{3}$}   (q_d)
    (q_d) edge [bend left]  node {$\mu_{4}$}       (q_4)
          edge [bend left]  node {$\lambda_{M-2}$} (q_5)

    (q_5) edge [bend left]  node {$\mu_{M-1}$}     (q_d)
          edge [bend left]  node {$\lambda_{M-1}$} (q_6)
          edge [loop above] node { }               (   )
    (q_6) edge [bend left]  node {$\mu_{M}$}       (q_5)
          edge [loop above] node { }               (   );
\end{tikzpicture}

\begin{figure}[ht]
  \centering
  \begin{subfigure}[b]{0.4\linewidth}
    \includegraphics[width=\linewidth]{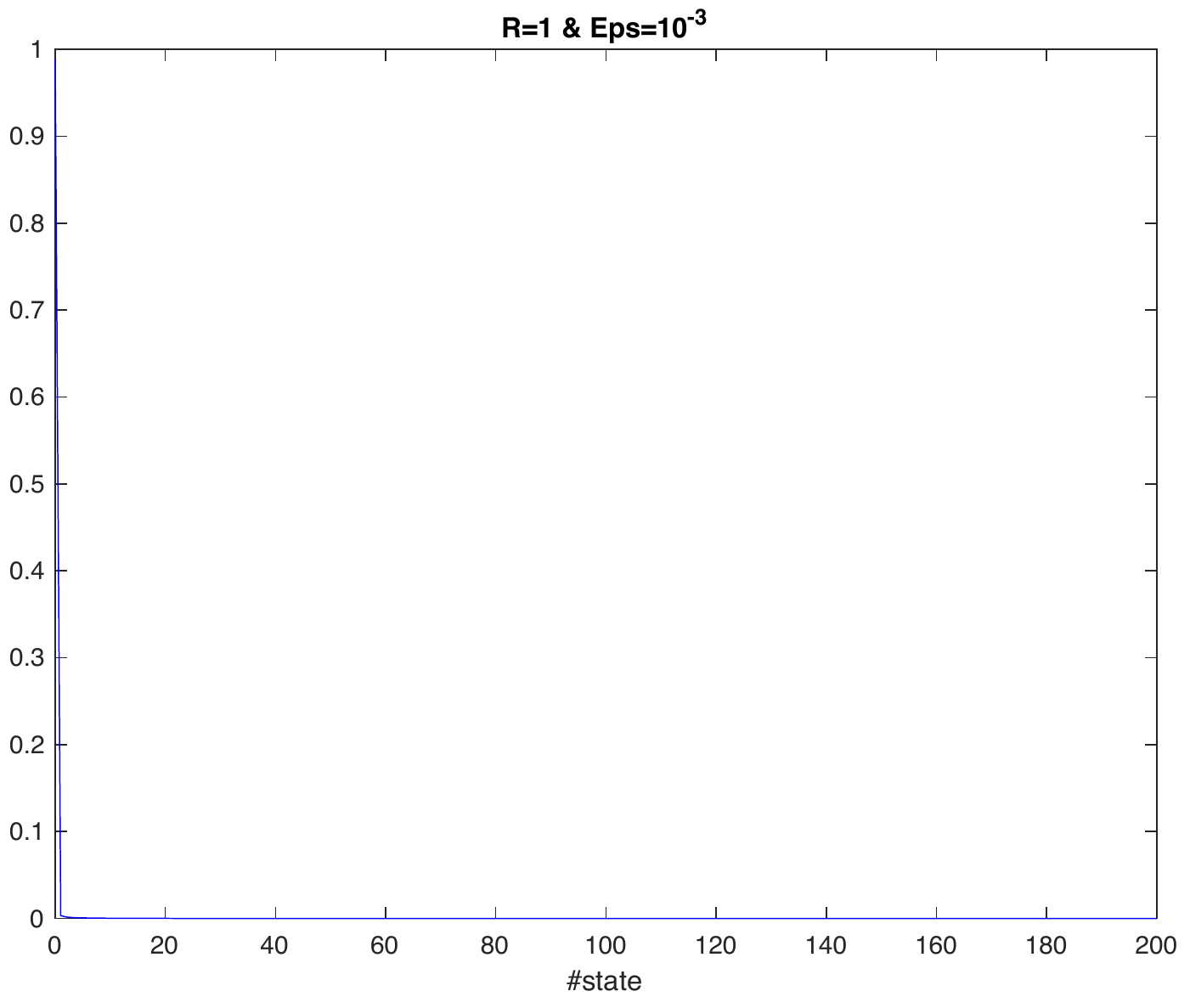}
     \caption{ Distribution concentrates at $0$}
         \label{fig31}
  \end{subfigure}
  \begin{subfigure}[b]{0.4\linewidth}
    \includegraphics[width=\linewidth]{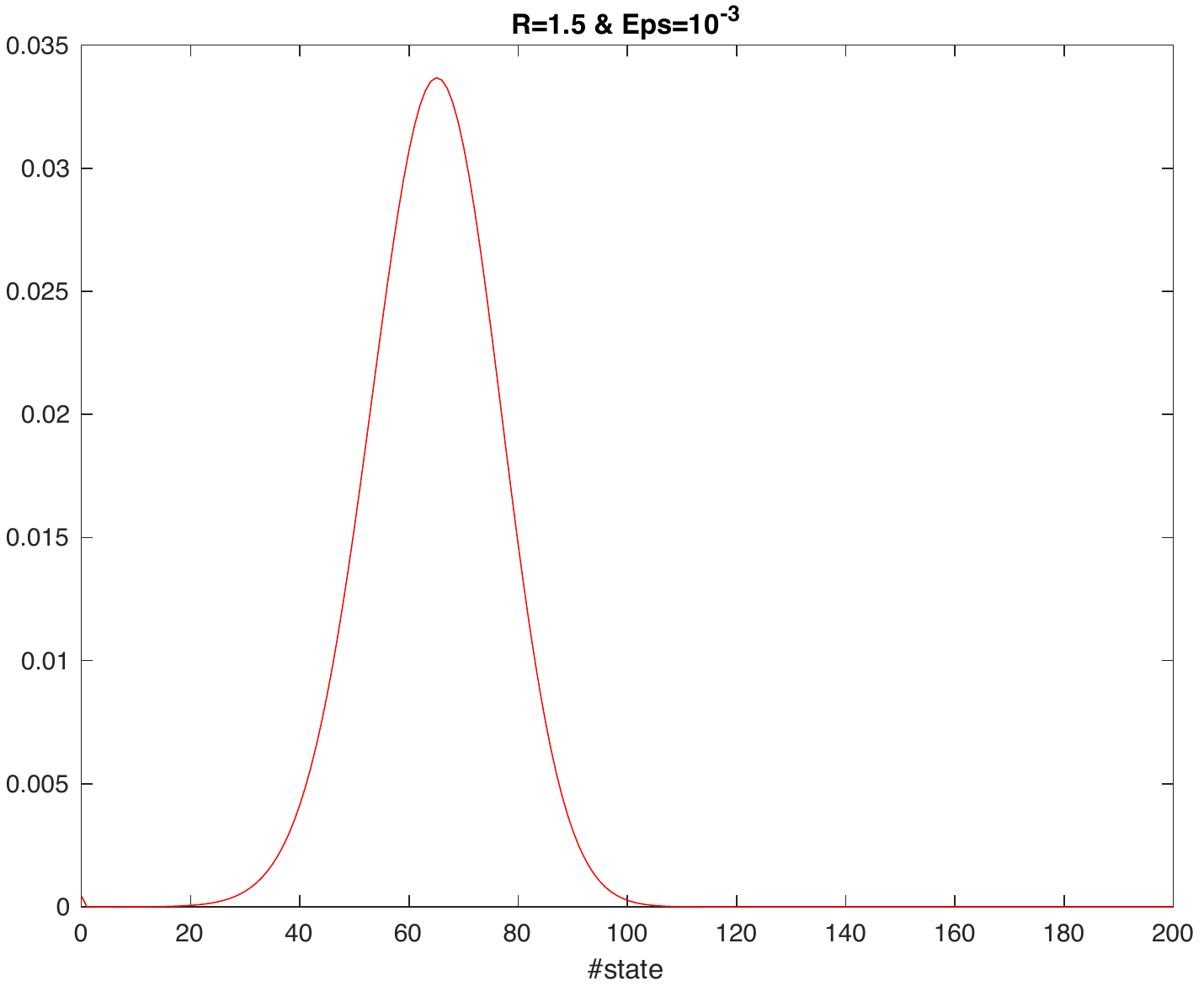}
      \caption{Bell-shaped distribution around 66}
         \label{fig32}
  \end{subfigure}\\
  \begin{subfigure}[b]{0.4\linewidth}
    \includegraphics[width=\linewidth]{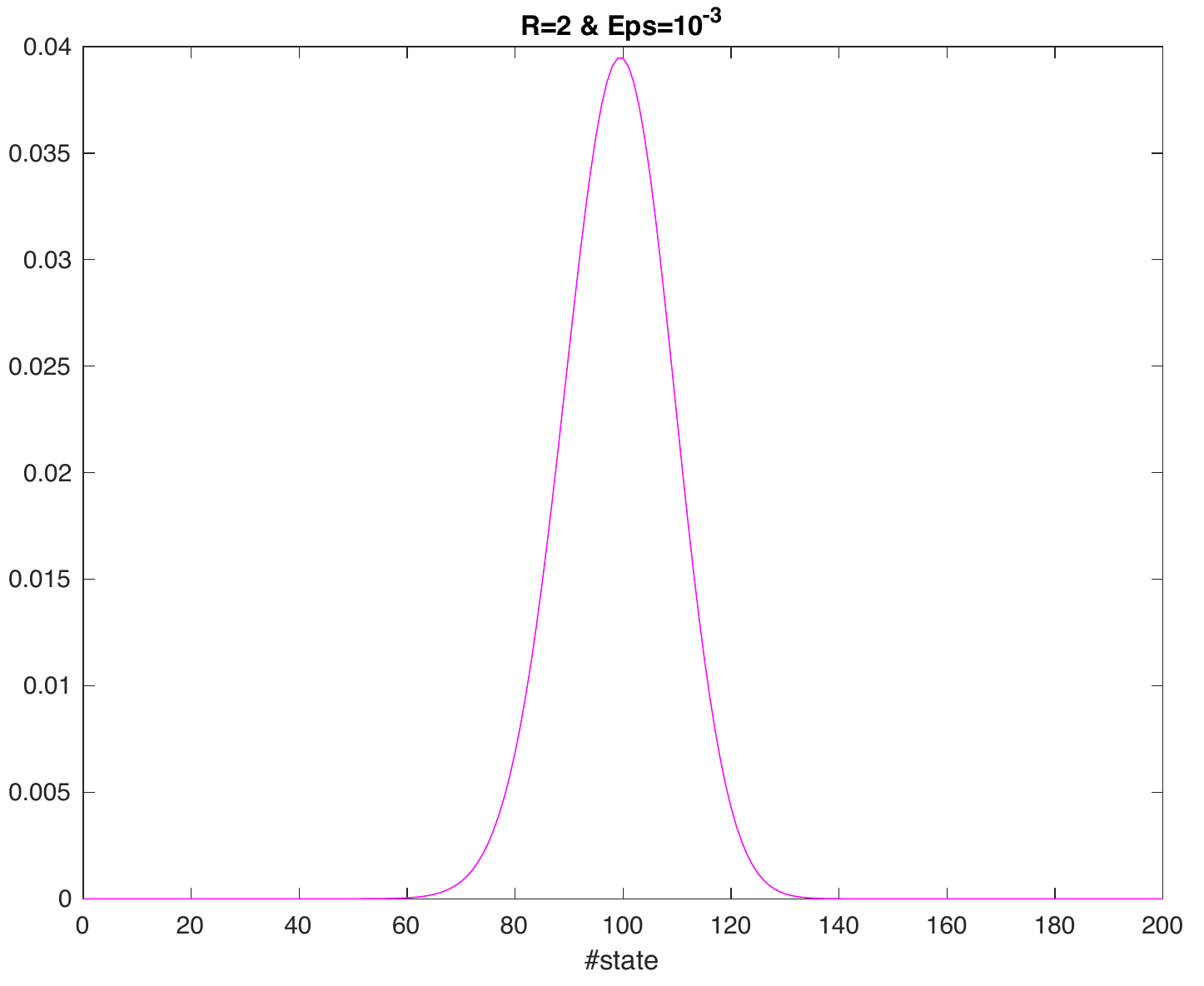}
    \caption{Bell-shaped distribution around 100}
         \label{fig33}
  \end{subfigure}
  \begin{subfigure}[b]{0.4\linewidth}
    \includegraphics[width=\linewidth]{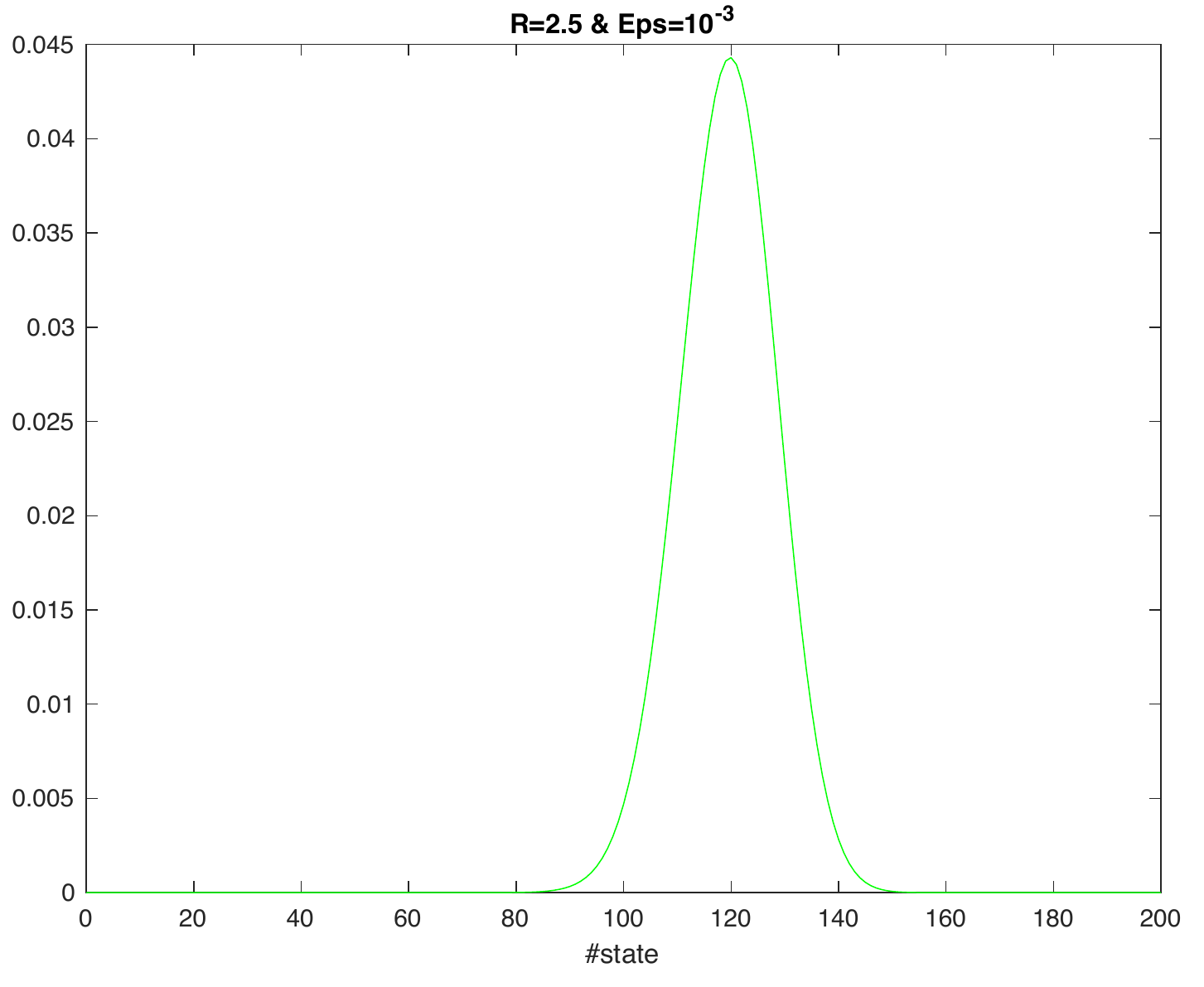}
     \caption{Bell-shaped distribution around 120}
         \label{fig34}
  \end{subfigure}
  \caption{Distribution ($P_{M,\epsilon}$) : $M=100$, $\alpha=0.3$, $\beta=R \alpha$}
  \label{fig3}
\end{figure}

\noindent
However, $\lambda_0 =\beta\frac{(M-0)} {M}0=0$ and according to the stochastic process theory, we have two equivalence classes: \{1, 2, ...., M\} and \{0\}. In order to have an irreducible Markov Process, we introduce an external source of disease through a small nonnegative parameter ($\epsilon>0$) in the infection transition rate. We have a new Process $P_ {M, \epsilon} $ on the state space $\{0,1,\ldots,M\}$.
 \begin{align*}
   \lambda_i^{[\epsilon]} =\beta\frac{(M-i)} {M}i+\epsilon \hspace{5mm} (i\rightarrow i+1) \quad  \quad   \mu_i=\alpha i \hspace{5mm} (i\rightarrow i-1)
   \end{align*} 
According to Karlin et al\cite {karlin1975}, we define the following quantities:
  \begin{align*}
   \theta_0= 1    \quad \quad \theta_i^{[\epsilon]} =\frac{\lambda_0^{[\epsilon]} \lambda_1^{[\epsilon]} \cdots \lambda_{i-1} ^{[\epsilon]}} {\mu_1\mu_2\cdots \mu_i} \hspace{5mm}
   \hbox{for $i=1,\ldots,M$}.
\end{align*} 
And therefore, we have the stationary distribution of $P_ {M, \epsilon} $\cite{karlin1975}.
\[ \pi_{M}^{[\epsilon]}(i)=
 \begin{cases}
 \frac{\theta_0}{\theta_0 + \sum_{j=1}^M\theta_j^{[\epsilon]}} & \quad i=0\\
\frac{\theta_i^{[\epsilon]}}{\theta_0 + \sum_{j=1}^M\theta_j^{[\epsilon]}} & \quad (i=1,2,\dots,M)
 \end{cases}
\]
In order to appreciate the shape of the stationary distribution, we look at four cases with Reproduction Number (R): $R=1$, $R=1.5$, $R=2$ and $R=2.5$. For small $\epsilon=10^ {-3} $, $M=200$, and $R=1$, the distribution is mostly concentrated at state $0$, but not only at state $0$ as shown in Fig \ref{fig31}. 

\noindent
When $R$ greater than 1, the distribution is continuous and well-spread. For $R=1.5$, the distribution has a bell-shape and spreads on the left side around 66 infections in Fig \ref{fig32}. For $R=2$, as shown in Fig \ref{fig33}, the distribution is still a bell-shape, but the spread is centered at 100 infections.\\

\begin{figure}[ht]
  \centering
  \begin{subfigure}[b]{0.35\linewidth}
    \includegraphics[width=\linewidth] {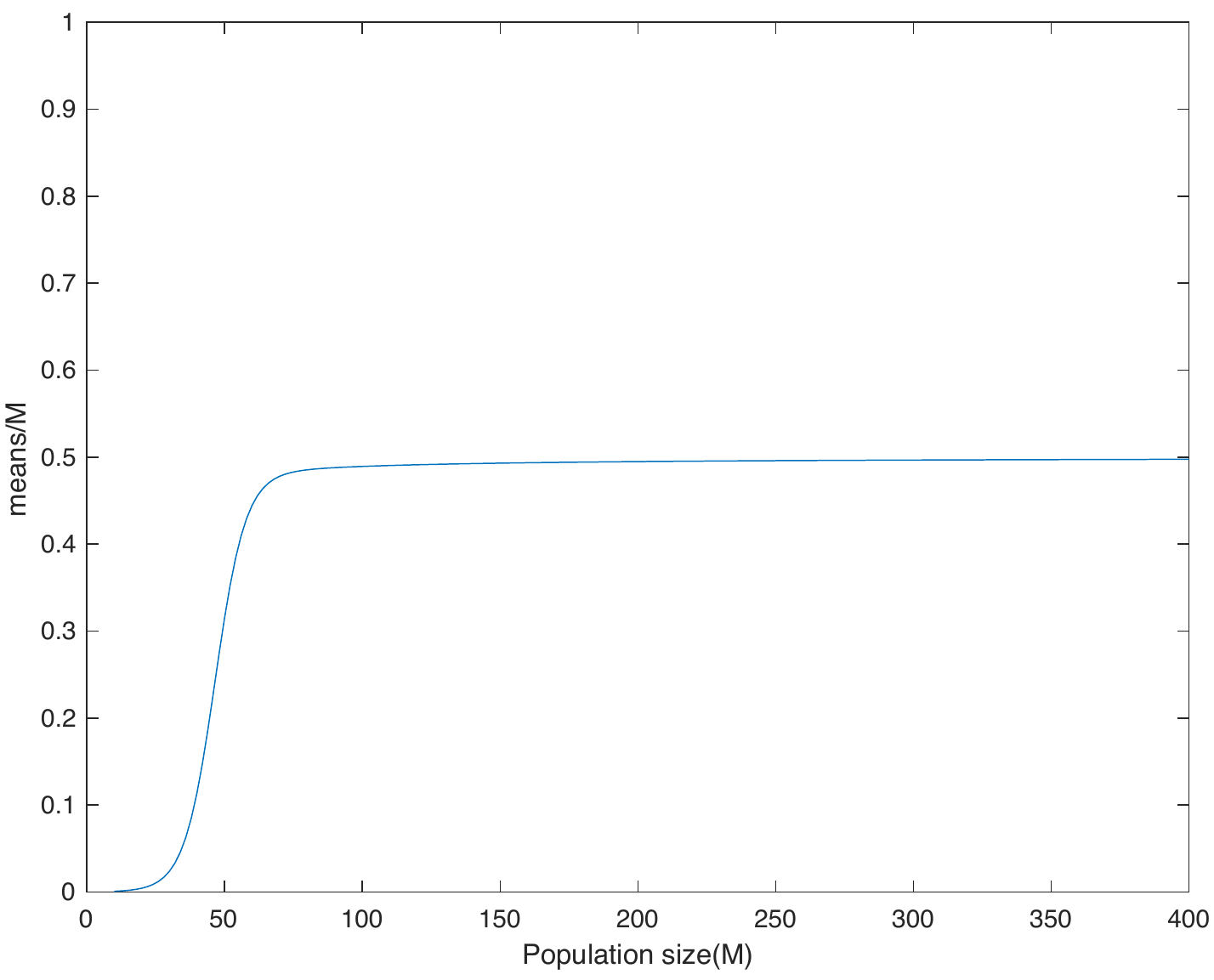}
     \caption {Mean/M converges to $1-\frac{1}{R} $}
         \label{fig41}
  \end{subfigure}
  \begin{subfigure}[b]{0.35\linewidth}
    \includegraphics[width=\linewidth] {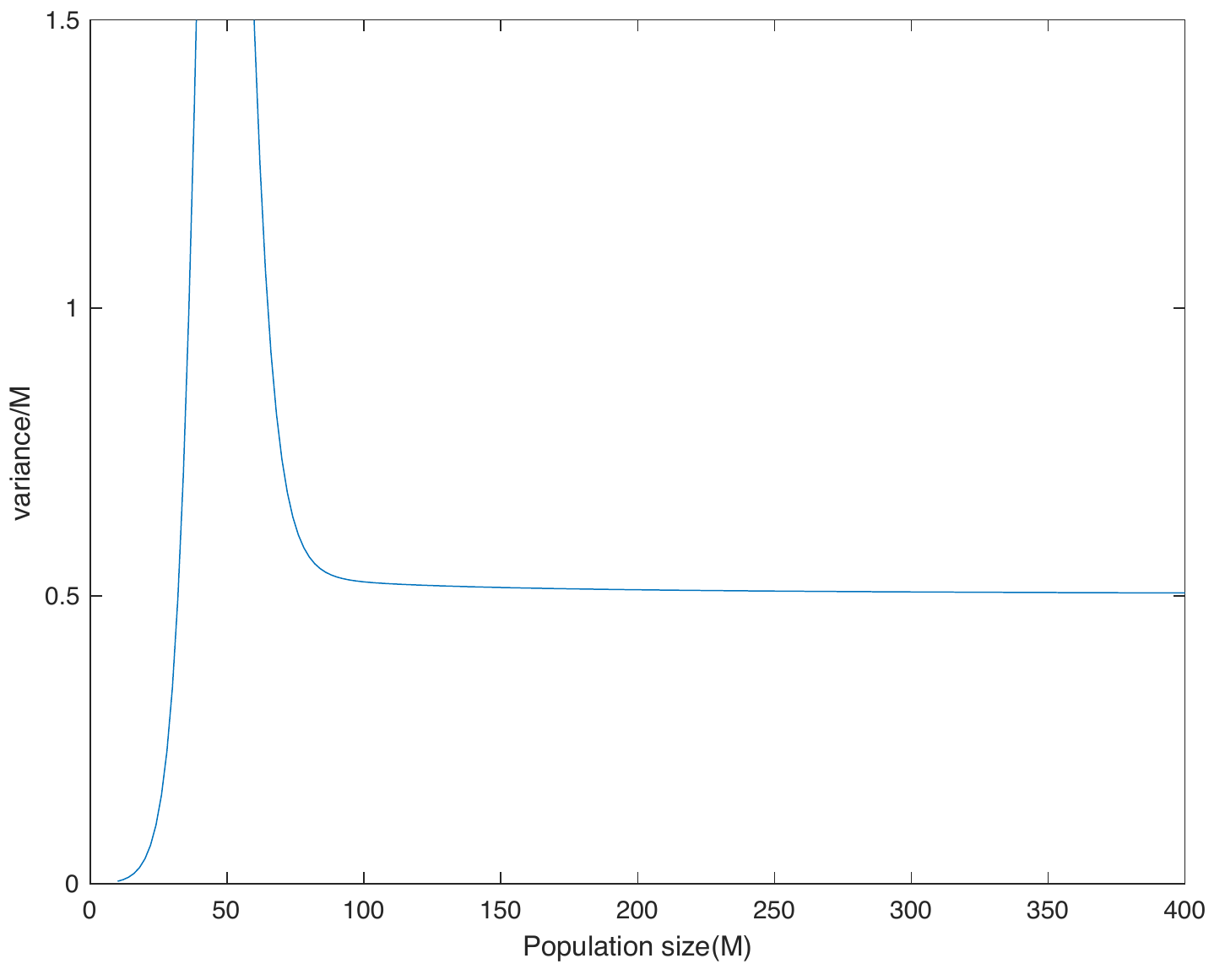}
      \caption {Variance/M converges to $\frac{1}{R} $}
         \label{fig42}
  \end{subfigure}\\
  \begin{subfigure}[b]{0.35\linewidth}
    \includegraphics[width=\linewidth] {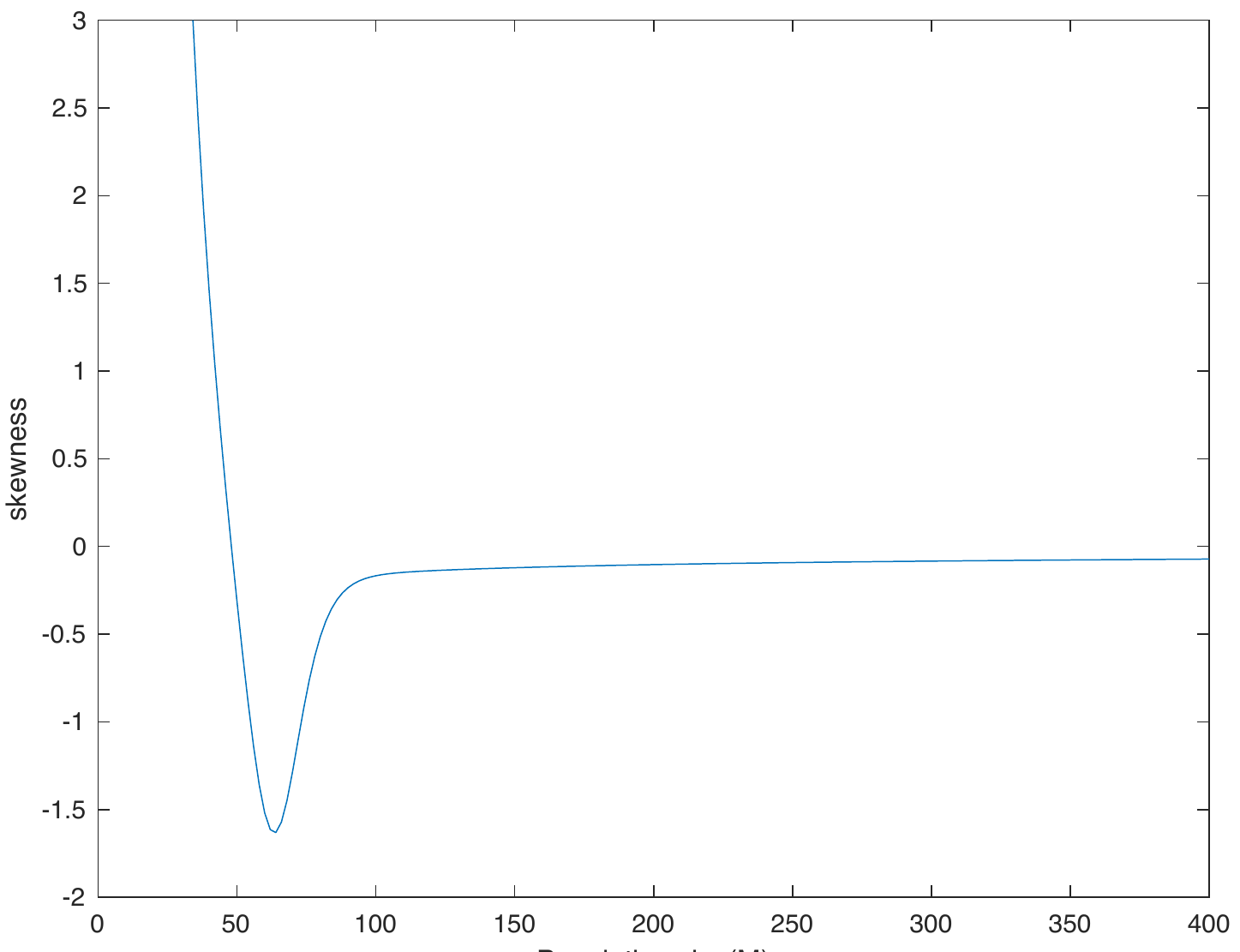}
    \caption {Skewness converges to $0$}
         \label{fig43}
  \end{subfigure}
  \begin{subfigure}[b]{0.35\linewidth}
    \includegraphics[width=\linewidth] {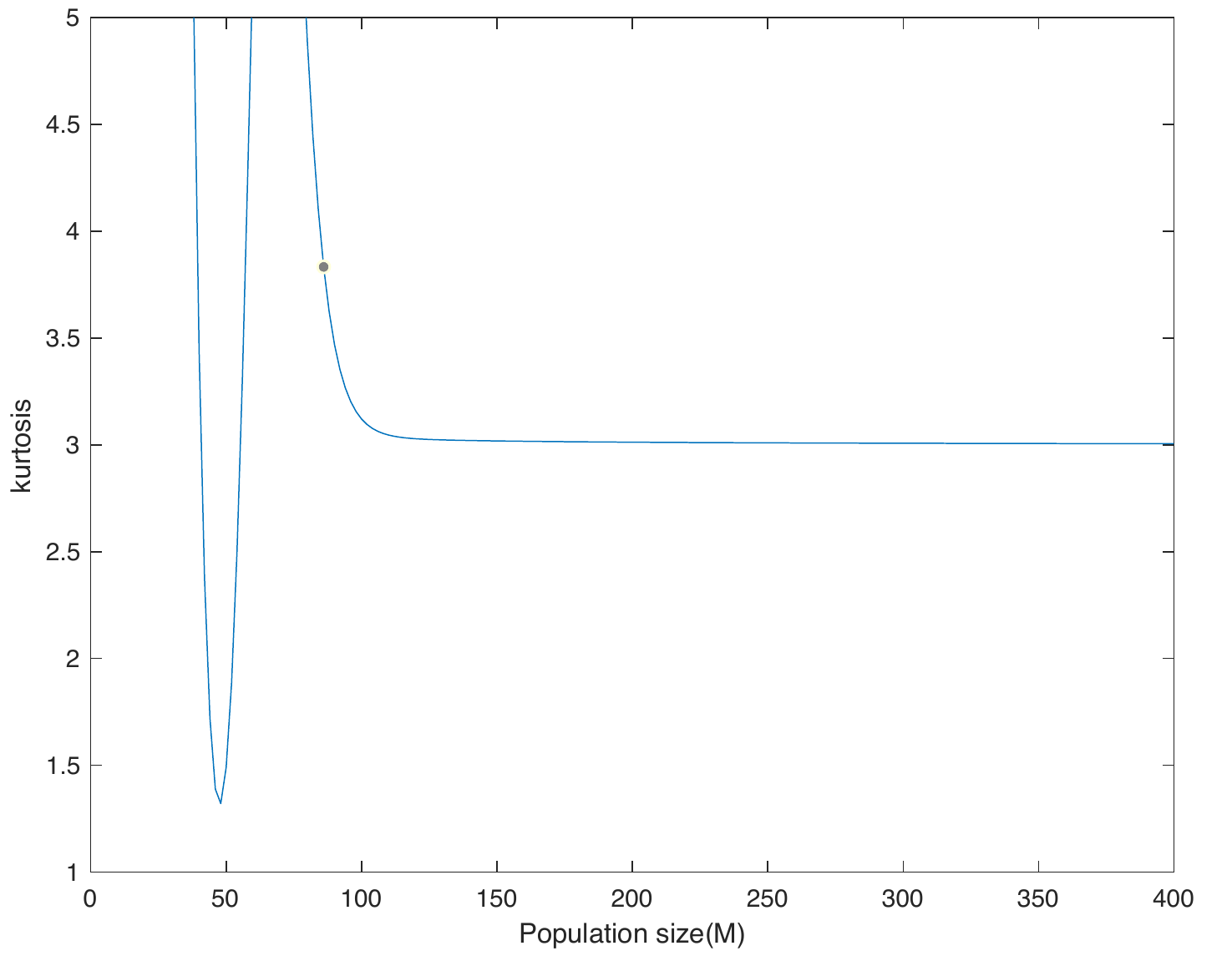}
     \caption {Kurtosis converges to $3$}
         \label{fig44}
  \end{subfigure}
  \caption {Central and dispersion characteristics of the distribution ($P_{M, \epsilon}$) : $R=2$, $\epsilon=10^{-4}$ and $\alpha=0.3$}
  \label{fig4}
\end{figure}

\noindent
Given the distribution ($P_ {M, \epsilon} $), we can look at some measures of central tendency and dispersion. In Fig \ref{fig4}, four statistical indicators (mean, variance, skewness and kurtosis) are used to summarize the characteristics of the distribution. As illustrated in Fig \ref{fig41}, the mean\footnote{$\bar{x}=\sum_{j=0} ^{M} \pi_{M}^{[\epsilon]} (j)j$} of infected size is a function of the population size ($M$). In fact, when the population size ($M$) is small ($M<100$), the mean increases at a growing rate before reaching a stable position as $M$ becomes greater than $100$. At this stage, the mean continues to increase but at a constant rate ($1-\frac{1}{R} $). In Fig \ref{fig42}, the variance\footnote{$\sigma^2=\sum_{j=0} ^{M} {\pi_{M}^{[\epsilon]} (j)(j-\bar{x}) ^2} $} is also a function of the population size ($M$) and is unstable when $M$ is not large ($M<100$). When $M$ becomes large ($M>100$), the variance reaches a stable phase where the increasing is at a constant rate ($\frac{1}{R} $).

\noindent
As shown in Fig \ref{fig4}, the results of the Skewness and Kurtosis indicators are independent of reproduction number ($R$).
The Skewness\footnote{$Sk=\sum_{j=0} ^{M} {\pi_{M}^{[\epsilon]} (j)(\frac{j-\bar{x}} {\sigma}) ^3} $} is an indicator of lack of symmetry, that is, both left and right sides of the distribution ($P_ {M, \epsilon} $) are unequal with respect to the mean. In Fig \ref{fig43}, the Skewness as a function of $M$ shows that the distribution of the system lacks symmetry when $M$ is not large enough ($M<100$). When $M$ become large enough, the Skewness converges to $0$; and the symmetric natures of the distribution appears. \\
The Kurtosis\footnote{$K=\sum_{j=0} ^{M} {\pi_{M}^{[\epsilon]} (j)(\frac{j-\bar{x}} {\sigma}) ^4} $} is a measure of how heavy-tailed or light-tailed the distribution ($P_ {M, \epsilon} $) is relative to a normal distribution. In Fig \ref{fig44}, the Kurtosis as a function of $M$ shows that the SIS model alternates between heavy-tailed and light-tailed before reaching a stable value of $3$, when $M$ becomes large enough. It is important to point out that the normal distribution has kurtosis equal to 3.\\
 Figs \ref{fig3} and \ref{fig4} provide evidence that the infected size follows a normal distribution when the population size(M) reaches a certain threshold.

\section{Infected Size Distribution: Analytical Results}
\noindent
We revise the continuous-time Markov chain on the state space $\{0,1,\ldots,M-1\}$ with transition rates. 
\begingroup
\setlength{\abovedisplayskip}{0pt}
\setlength{\belowdisplayskip}{0pt}
 \begin{align*}
\lambda_k &= \left\{\begin{array}{ll}
   \beta\left(\frac{M-k}{M}\right)k &\mbox{for $k=1\ldots M-1$} \\
    \epsilon & \mbox{for $k=0$}
  \end{array}\right. \hspace{15mm}(k\rightarrow k+1)  \\
      \mu_k & = \; \alpha k \hspace{15mm} (k\rightarrow k-1)
\end{align*} 
\endgroup

\noindent
In this revision version, the external factor $\epsilon > 0$ is kept only for $\lambda_0$. Here $\alpha$ and $\beta$ are strictly positive parameters, and $\epsilon$ is a non-negative parameter.
We define $\theta_0,\theta_1^{[\epsilon]},\ldots,\theta_M^{[\epsilon]}$ by
 \begin{align*}
   \theta_0 & = 1   \\
   \theta_i^{[\epsilon]}  & =\frac{\lambda_0 \lambda_1 \cdots \lambda_{i-1}}{\mu_1\mu_2\cdots \mu_i} \hspace{5mm}
   \hbox{for $i=1,\ldots,M$}.
\end{align*}
The equilibrium distribution of $P_{M,\epsilon}$ is derived as follows:
\[ \pi_{M}^{[\epsilon]}(i)=
 \begin{cases}
 \frac{\theta_0}{\theta_0 + \sum_{j=0}^{M-1}\theta_{(M-j)}^{[\epsilon]}} & \quad i=0\\
\frac{\theta_i^{[\epsilon]}}{\theta_0 + \sum_{j=0}^{M-1}\theta_{(M-j)}^{[\epsilon]}} & \quad (i=1,2,\dots,M)
 \end{cases}
\]
 \begin{lemma}\label{lem1} Assume $\beta>0$, $\alpha>0$, $\epsilon>0$, $M>0$ and $R=\frac{\beta}{\alpha}$. We have :
 \begin{align*}
   \theta_0 & = 1   \\
   \theta_{M-k}^{[\epsilon]} & = \frac{\epsilon}{R\alpha(M-k)}\left(\frac{R}{M}\right)^{M-k}\frac{M!}{k!}  \hspace{5mm} \hbox{for $k=0,\ldots,M-1$}
\end{align*}
\end{lemma} 
 \noindent
\textbf{Proof:}\\ 
  \begin{align*}
  \theta_{M-k}^{[\epsilon]} & =\frac{\lambda_0\lambda_1 \cdots\lambda_{(M-k-1)}}{\mu_1\mu_2\cdots \mu_{(M-k)}}=\frac{\lambda_0}{\mu_{(M-k)}}\prod_{j=1}^{M-k-1}\frac{\lambda_j}{\mu_j}=\frac{\epsilon}{\alpha(M-k)}\prod_{j=1}^{M-k-1}{[\frac{\beta}{\alpha}{\left(\frac{M-j}{M}\right)}]} \\
& =\frac{\epsilon}{\alpha(M-k)}\prod_{j=1}^{M-k-1}{R{\left(\frac{M-j}{M}\right)}}= \frac{\epsilon}{R\alpha(M-k)}\left(\frac{R}{M}\right)^{M-k}\frac{M!}{k!}\hspace{10mm} \hbox{$\Box$}
\end{align*} 
 \subsection {Properties of Poisson Distribution}
 Some properties\cite{jeremy1963} of Poisson distribution will be stated with proof and the results will be applied in the next subsection.
 
 \begin{lemma}
 \label{lem2} Suppose $X$ follows a Poisson distribution with parameter $\lambda$ and \\
 $\mu(d)= E[X|X\leq d] \hspace{2mm} \hbox{ $\forall d\in \N^* $ } $ \ \ 
Then:
 \begin{align*}
 \mu(d)=\lambda \frac{g(d-1)}{g(d)} \hspace{10mm}\ \ where \ \ g(d)=\sum_{i=0}^{d} {\frac{\lambda^i}{i!}}\ \ and \ \
\lim_{d\to\infty}\frac{g(d-1)}{g(d)}=1 
  \end{align*}
\end{lemma} 

 \noindent
\textbf{Proof:}\\
 Let us define the following function \ \ $p(x,\lambda,d)=P(X=x| X\leq d)$ \ \ for $x=0,\ldots,d$
   \begin{align*}
p(x,\lambda,d)=P(X=x| X\leq d)=\frac{p(X=x)}{p(X\leq d)}=\frac{\frac{\lambda^x}{x!}}{\sum_{i=0}^{d} {\frac{\lambda^i}{i!}}}=\frac{\frac{\lambda^x}{x!}}{g(d)}=\frac{g(d-1)}{g(d)} p(x,\lambda,d-1)
  \end{align*}
 with 
$\frac{g(d-1)}{g(d)} =\frac{\sum_{i=0}^{d-1} {\frac{\lambda^i}{i!}}}{\sum_{i=0}^{d} {\frac{\lambda^i}{i!}}}=1-\frac{\frac{\lambda^d}{d!}}{\sum_{i=0}^{d} {\frac{\lambda^i}{i!}}}$\ \ and \ \
$\lim_{d\to\infty} \frac{g(d-1)} {g(d)} =1-\frac{ \lim_{d\to\infty}\frac{\lambda^d}{d!}}{ \lim_{d\to\infty}\sum_{i=0}^{d} {\frac{\lambda^i}{i!}}} =1$\\
We also have   \ \ 
$\lambda\frac{dg}{d\lambda} (d)=\sum_{j=1}^{d} {j\frac{\lambda^{j}}{j!}}=\lambda g(d-1)$ \ \ and the result follows\\
   \begin{align*}
 \mu(d)= E[X|X\leq d] =\sum_{j=0}^{d} {jp(j,\lambda,d)}=\frac{\sum_{j=1}^{d} {j\frac{\lambda^{j}}{j!}}}{g(d)}=\frac{\lambda\frac{dg}{d\lambda}(d)}{g(d)}=\lambda \frac{g(d-1)}{g(d)} \hspace{10mm} \hbox{$\Box$}
   \end{align*}

 \begin{corollary}\label{cor1} Assume $R>1$, $M>0$, and $X$ follows a Poisson distribution with parameter $\frac{M}{R}$.
Then:
 $E[{\frac{X}{M}|X<M}] =\frac{1}{R}\frac{g(M-2)}{g(M-1)}$ \ \ $\text{with}$ \ \ $\lim_{M\to\infty}\frac{g(M-2)}{g(M-1)}=1$ 
 \end{corollary}
 \noindent
\textbf{Proof:}\\ 
From lemma \ref{lem2}, $\lambda=\frac{M}{R}$ and $d=M-1$; 
\begingroup
\setlength{\abovedisplayskip}{0pt}
\setlength{\belowdisplayskip}{0pt}
    \begin{align*}
 E[\frac{X}{M}|X<M] = E[\frac{X}{M}|X\leq M-1] = \frac{1}{M}E[X|X\leq M-1]=\frac{1}{M}\mu(M-1)=\frac{1}{R}\frac{g(M-2)}{g(M-1)}     
     \end{align*}
     \endgroup
  and \ \ $E[\frac{X}{M}|X<M] =\frac{1}{R}\frac{g(M-2)}{g(M-1)}$ $\Box$\\
 \begin{lemma}\label{lem3}
  Assume $X$ follows a Poisson distribution with parameter $\lambda$ and let $a>\lambda$.\\
 We have: $P({X>a})\leq \mathrm{e}^{-\lambda + a - a{\log(\frac{a}{\lambda})}}$
  \end{lemma} 
 \noindent
\textbf{Proof:}\\
 $M(\theta)= \mathrm{e}^{\lambda(\mathrm{e}^{\theta} -1)}$ is the moment generating function of the Poisson distribution. \\
 $M(\theta) = E(\mathrm{e}^{\theta X})= \sum_{i=0}^{k\to\infty}{\mathrm{e}^{k\theta}P(X=k)}>\mathrm{e}^{a\theta}P(X>a)$ and
 $P(X>a)<\mathrm{e}^{\lambda(\mathrm{e}^{\theta} -1) -a\theta}$ $\forall \theta\in \R$\\
Therefore, $ P(X>a)\leq \inf_{\theta\in \R}{\{\mathrm{e}^{\lambda(\mathrm{e}^{\theta} -1) -a\theta}\}}= \mathrm{e}^{-\lambda + a - a{\log(\frac{a}{\lambda})}}$.
The function $\psi(\theta)=\mathrm{e}^{\lambda(\mathrm{e}^{\theta} -1) -a\theta}$ reaches its minimum at $\theta^*=\log(\frac{a}{\lambda})$ $\Box$\\
 \begin{lemma}\label{lem4} 
 Assume $X$ follows a Poisson distribution with parameter $\lambda=\frac{M}{R}$.\\
Then $P({X>\frac{M}{R} + \delta M})\leq \mathrm{e}^{\frac{\phi(\delta R)}{R}M}$ \ \ $\forall \delta>0$\ \ where \ \  $\phi()$ is a function and $ \phi(\delta R)<0$   
  \end{lemma} 
 \noindent
\textbf{Proof:}\\
 Let us define $\phi(x)=x-(1+x) \log(1+x)$ and it can be shown that : $ \phi(x)< 0$ \ \ $\forall x>0$ \\
For $x=\delta R$, we have $ \phi(\delta R)=\delta R-(1+\delta R)\log(1+\delta R) < 0 $\\ 
 For $a=\frac{M}{R}+\delta M$ and $\lambda=\frac{M}{R}$, we apply Lemma \ref{lem3}
  \begin{align*}
 P({X>\frac{M}{R}+\delta M})&\leq \mathrm{e}^{-\lambda + \frac{M}{R}+\delta M - (\frac{M}{R}+\delta M){\log(\frac{\frac{M}{R}+\delta M}{\lambda})}} = \mathrm{e}^{-\frac{M}{R} + \frac{M}{R}+\delta M - (\frac{M}{R}+\delta M){\log(\frac{\frac{M}{R}+\delta M}{\frac{M}{R}})}}\\
 &\leq \mathrm{e}^{\frac{M}{R}(\delta R - (1+\delta R){\log(1+\delta R)})}\\
 &\leq \mathrm{e}^{\frac{\phi(\delta R)}{R}M} \hspace{10mm} \hbox{$\Box$} \end{align*}

 \subsection {Limit Superior and Limit Inferior of the Expectation} 
   
 \begin{lemma}\label{lem5} 
 Assume $R>1$, $M>0$, $X$ follows a Poisson distribution with parameter $\lambda=\frac{M}{R}$.\\ Then $\limsup_{M\to\infty} E{(\frac{M}{M-X}|X<M)}\leq\frac{1}{1-\frac{1}{R}}$
\end{lemma} 
\noindent
\textbf{Proof:}\\
 $I(X<M)$ is an indicator function, \\
\begin{align*}
E[{\frac{M}{M-X}I(X<M)}] = E[{\frac{M}{M-X}I(X\leq M*)}] + E[{\frac{M}{M-X}I(M*<X<M)}]
\end{align*}
where $M*=\frac{M}{R}+\delta M$ and $\delta>0$. \\ 
We assume that $M*<M$, which is equivalent to $0<\delta<\frac{R-1}{R}$\\
\begin{align}
E[{\frac{M}{M-X}I(X\leq M*)}]\leq \frac{1}{1-\frac{1}{R}-\delta}P(X\leq \frac{M}{R}+\delta M)
\end{align}
  \begin{align}
  	E[{\frac{M}{M-X}I(M*<X<M)}]\leq MP(\frac{M}{R}+\delta M<X)
  \end{align}
By applying Lemma \ref{lem4}, 
$E[{\frac{M}{M-X}I(M*<X<M)}] \leq M\mathrm{e}^{\frac{\phi(\delta R)}{R}M}$ \ \ where $\phi(\delta R)<0$.\\
 \begin{align*}
E[{\frac{M}{(M-X)}|X<M}] =\frac{E[{\frac{M}{M-X}I(X<M)}]}{P(X<M)}&\leq \frac{P(X\leq \frac{M}{R}+\delta M)}{P(X<M)} \frac{1}{1-\frac{1}{R}-\delta}+\frac{1}{P(X<M)} M\mathrm{e}^{\frac{\phi(\delta R)}{R}M} \\
&\leq \frac{1}{1-\frac{1}{R}-\delta}+\frac{1}{P(X<M)} M\mathrm{e}^{\frac{\phi(\delta R)}{R}M} 
\end{align*} 
$\lim_{M\to\infty}\frac{1}{P(X<M)} M\mathrm{e}^{\frac{\phi(\delta R)}{R}M} =0$ and we have:
    \begin{align*}
 \lim_{\delta\to 0} \limsup_{M\to\infty}E[{\frac{M}{(M-X)}|X<M}] &\leq \lim_{\delta\to 0} \frac{1}{1-\frac{1}{R}-\delta} \hspace{10mm}
   \hbox{for \ \ $0<\delta <\frac{R-1}{R}$}\\
 \limsup_{M\to\infty}E[{\frac{M}{(M-X)}|X<M}] &\leq \frac{1}{1-\frac{1}{R}}  \hspace{10mm}
   \hbox{$\Box$} \end{align*}

\begin{lemma}\label{lem5a} 
 Assume $R>1$, $M>0$, $X$ follows a Poisson distribution with parameter $\lambda=\frac{M}{R}$.\\ Then $\liminf_{M\to\infty} E{(\frac{M}{M-X}|X<M)}\geq\frac{1}{1-\frac{1}{R}}$
\end{lemma} 
\noindent
\textbf{Proof:}\\
The function $f_M(x)=\frac{M}{M-x}$ is convex over $ 0\leq x<M$. Using the Jensen Inequality property, 
 \begin{align}\label{fig01}f_M(E[{X|X<M}] )\leq E[{f_M(X)|X<M}]
 \end{align} 
From corollary \ref{cor1}, $E[{X|X<M}] =\frac{M}{R}\frac{g(M-2)}{g(M-1)}$\\ 
From (\ref{fig01}), we have 
\begin{align*}  
E[{\frac{M}{M-X}|X<M}]\geq \frac{M}{M-E[{X|X<M}]} = \frac{M}{M-\frac{M}{R}\left(\frac{g(M-2)}{g(M-1)} \right)}
\end{align*}
\noindent  
We take the limit
\begin{align*}  
\liminf_{M\to\infty} E[{\frac{M}{M-X}|X<M}]\geq \liminf_{M\to\infty}{\frac{1}{1-\frac{1}{R}\left(\frac{g(M-2)}{g(M-1)} \right)} }= \frac{1}{1-\frac{1}{R}}
\end{align*} 

From lemma \ref{lem5} and lemma \ref{lem5a}, we have: 
 \begin{align} 
 \label{fig02} 
\lim_{M\to\infty} {E[{\frac{M}{M-X}|X<M}]} =\frac{1}{1-\frac{1}{R}}\hspace{10mm} \hbox{$\Box$}
   \end{align}
 
 \subsection {Approximation of the Asymptotic Distribution of $P_{M,\epsilon}$} 
 
 \begin{lemma}\label{lem6}  
   Assume $\beta>0$, $\alpha>0$, $\epsilon>0$, $M>0$ with $R=\frac{\beta} {\alpha}>1$.
\begin{align*}
  \sum_{k=0}^{M}\theta_{(M-k)}^{[\epsilon]}=1+ \sum_{k=0}^{M-1}\theta_{(M-k)}^{[\epsilon]} \sim C(M){\frac{R}{R-1}}{\mathrm{e}^{\left(\frac{M}{R}\right)}} \hspace{20mm}
   \hbox{As $M\rightarrow\infty$ \ \ and \ \ $C(M)= \frac{M!\epsilon}{MR\alpha}\left(\frac{R}{M}\right)^{M} $ } 
    \end{align*}
\end{lemma}
\noindent
\textbf{Proof:}\\ 
 \begin{align*}
 \sum_{k=0}^{M-1}{\theta_{M-k}^{[\epsilon]}} &= \sum_{k=0}^{M-1}{\frac{\epsilon}{R\alpha(M-k)}\left(\frac{R}{M}\right)^{M-k}\frac{M!}{k!}} \hspace{10mm}
   \hbox{ ($\theta_{M-k}^{[\epsilon]}$ from Lemma \ref{lem1})} \\
 &=C(M){\mathrm{e}^{\left(\frac{M}{R}\right)}}\sum_{k=0}^{M-1}{\frac{M}{M-k}\frac{1}{k!}\left(\frac{M}{R}\right)^{k}{\mathrm{e}^{\left(-\frac{M}{R}\right)}}} \hspace{10mm}
   \hbox{with $C(M)= \frac{M!\epsilon}{MR\alpha}\left(\frac{R}{M}\right)^{M} $}\\
 &=C(M){\mathrm{e}^{\left(\frac{M}{R}\right)}}\sum_{k=0}^{M-1}{\frac{M}{M-k}P[X=k]} \hspace{10mm}
   \hbox{with $X \sim Poisson(\frac{M}{R})$} \\
 &=C(M){\mathrm{e}^{\left(\frac{M}{R}\right)}}E[{\frac{M}{M-X}I(X<M)}] \hspace{5mm}
   \hbox{($I(X<M)$ indicator function)} 
    \end{align*}  
From the conditional expectation:
   \begin{align}
    \label{eq:l01}
     E[{\frac{M}{(M-X)}I(X<M)}] =P[X<M]E[{\frac{M}{M-X}|X<M}] 
     \end{align} 
Previously, we show that  
 \begin{align*}
\sum_{k=0}^{M-1}{\theta_{(M-k)}^{[\epsilon]}} = C(M){\mathrm{e}^{\left(\frac{M}{R}\right)}}E[{\frac{M}{M-X}I(X<M)}]
\end{align*}
We deduce the following relation
 \begin{align}
 \label{eq:l02}
 \frac{1+ \sum_{k=0}^{M-1}{\theta_{(M-k)}^{[\epsilon]}}}{C(M){\mathrm{e}^{\left(\frac{M}{R}\right)}(\frac{R}{R-1})}} &= (1-\frac{1}{R})E[{\frac{M}{M-X}I(X<M)}] +\frac{1}{C(M){\mathrm{e}^{\left(\frac{M}{R}\right)}(\frac{R}{R-1})}}
      \end{align}
\noindent
From the results (\ref{eq:l01}) and (\ref{fig02}), we have
 \begin{align*}
\lim_{M\to\infty}{(1-\frac{1}{R})E[{\frac{M}{M-X}I(X<M)}]} =1
\end{align*} 
We need to show that \cite {Nzokem2020EpidemicDA,aubain2020}
   \begin{align}
    \label{eq:l1}
    \lim_{M\to\infty}{\frac{1}{C(M){\mathrm{e}^{\left(\frac{M}{R}\right)}(\frac{R}{R-1})}}} =0 \hspace{10mm} \hbox{ $\forall R>1 $} 
 \end{align}
From (\ref{eq:l02}) and (\ref{eq:l1}), we have : 
\begin{align*}
	\lim_{M\to\infty}{ \frac{1+ \sum_{k=0}^{M-1}{\theta_{(M-k)}^{[\epsilon]}}}{C(M){\mathrm{e}^{\left(\frac{M}{R}\right)}(\frac{R}{R-1})}}} =1 \hspace{10mm} \hbox{$\Box$}
\end{align*}
 \begin{theorem}\label{lem7} 
  Assume $\beta>0$, $\alpha>0$, $\epsilon>0$, $M>0$ with $R=\frac{\beta}{\alpha}>1$.
 The infected size has the following equilibrium distribution.
\begin{align*}
  \pi_{M}^{[\epsilon]}(k) \sim \pi(k) \hspace{20mm}
   \hbox{As $M\rightarrow\infty$}
  \end{align*}
with 
\begin{align*}
    \pi(k) &={\frac{R-1}{R}}\frac{M}{(M-k)!k}\left(\frac{M}{R}\right)^{M-k}{\mathrm{e}^{\left(-\frac{M}{R}\right)}} \hspace{6mm}(k=1,2,\dots,M)   
\end{align*}
\end{theorem}
 \noindent
\textbf{Proof:}\\ 
For $k=1,2,\dots,M$ 
 \begin{align*}
   \pi_{M}^{[\epsilon]}(k) &= \frac{\theta_k^{[\epsilon]}}{1+ \sum_{j=0}^{M-1}\theta_{(M-j)}^{[\epsilon]}} = \frac{1}{\frac{1+ \sum_{j=0}^{M-1}\theta_{(M-j)}^{[\epsilon]}}{C(M)(\frac{R}{R-1}){\mathrm{e}^{\left(\frac{M}{R}\right)}}}}\frac{\theta_k^{[\epsilon]}}{C(M){(\frac{R}{R-1})}{\mathrm{e}^{\left(\frac{M}{R}\right)}}} \\
               &= \frac{1}{\frac{1+ \sum_{j=0}^{M-1}\theta_{(M-j)}^{[\epsilon]}}{C(M)(\frac{R}{R-1}){\mathrm{e}^{\left(\frac{M}{R}\right)}}}} \frac{R-1}{R}\frac{M}{(M-k)!k}{\left(\frac{M}{R}\right)^{M-k}}{\mathrm{e}^{-\left(\frac{M}{R}\right)}} \\
               &= \frac{1}{\frac{1+ \sum_{j=0}^{M-1}\theta_{(M-j)}^{[\epsilon]}}{C(M)(\frac{R}{R-1}){\mathrm{e}^{\left(\frac{M}{R}\right)}}}} \pi(k) 
  \end{align*} 
  We have the following quotient:
  \begin{align*}
  \frac{\pi_{M}^{[\epsilon]}(k) }{\pi(k) } &= \frac{1}{\frac{1+ \sum_{j=0}^{M-1}\theta_{(M-j)}^{[\epsilon]}}{C(M)(\frac{R}{R-1}){\mathrm{e}^{\left(\frac{M}{R}\right)}}}}
    \end{align*} 
The result follows from lemma \ref{lem6} $\Box$ \\

 \begin{theorem}\label{lem8} 
  Assume $\beta>0$, $\alpha>0$, $M>>1$ with $R=\frac{\beta}{\alpha}>1$. 
 The infected size follows asymptotically\footnote{As $M\rightarrow\infty$} a normal distribution with mean $\mu=(1-\frac{1}{R})M$ and variance $\sigma^2=\frac{M}{R}$.
\end{theorem} 
 \noindent
 \textbf{Proof:} \\
For k fixed, We know that $(M- k)! \sim \sqrt{2\pi}\mathrm{e}^{-(M-k)}(M-k)^{(M-k)+\frac{1}{2}}$  for  $M\rightarrow\infty$ \\
 
\noindent
We have the following equivalence when $M\rightarrow\infty$    \begin{align*} 
    \pi(k) = \frac{R-1}{R}\frac{M}{(M-k)!k}{\left(\frac{M}{R}\right)^{M-k}}{\mathrm{e}^{-\left(\frac{M}{R}\right)}} &\sim  \frac{R-1}{R\sqrt{2\pi}}\frac{M}{k(M-k)^{\frac{1}{2}}}{\left(\frac{M}{(M-k)R}\right)^{M-k}}{\mathrm{e}^{-\left(\frac{M}{R}\right)+(M-k)}} \\
    &= \psi(k)    
     \end{align*} 
     And we have 
\begin{align} 
\label{eq:99}
 \lim_{M\to\infty}{ \frac{\psi(k)}{\pi(k)}}  = 1
 \end{align} 
\noindent
  Supposed $x=k= (1-\frac{1}{R})M(1+\delta)$. By applying the second order Taylor's expansion techniques, we have the following expression \cite{aubain2020, Nzokem2020EpidemicDA}
     
 \begin{align*}
 \log(\psi(k)) & =\log(\frac{R-1}{R\sqrt{2\pi}}) + \log(\frac{M}{k}) - \frac{1}{2}\log(M-k) + (M-k)\log(\frac{M}{(M-k)R}) - \frac{M}{R} + (M-k) \\
& = - \frac{1}{2}\log(2\pi\frac{M}{R}) - \frac{1}{2}\frac{(R-1)^2}{R}M\delta^2 +[\frac{(R-3)}{2}\delta + \frac{1}{4}((R-1)^2+2)\delta^2 + O_1(\delta^3)] \nonumber \\
   &\qquad {}+[- \frac{1}{2}\frac{(R-1)^3}{R}M\delta^3 - (\frac{1-(R-1)\delta}{R} M+\frac{1}{2})O_2(\delta^3)]
\end{align*}
\noindent
We have a simple expression 

   \begin{align}
 \label{eq:100}
 \psi(k) &=\frac{1}{\sqrt{2\pi\frac{M}{R}}}\mathrm{e}^{\left\{-\frac{1}{2}\frac{(R-1)^2}{R}M\delta^2\right\}} \mathrm{e}^{\left\{\frac{(R-3)}{2}\delta + \frac{1}{4}((R-1)^2+2)\delta^2 + O_1(\delta^3)\right\}} \mathrm{e}^{\left\{- \frac{1}{2}\frac{(R-1)^3}{R}M\delta^3 - (\frac{1-(R-1)\delta}{R} M+\frac{1}{2})O_2(\delta^3)\right\}} 
 \end{align}
 \noindent
   Now, we can prove the Local Central Limit Theorem.  For $k$ integer, to get a convenient limit, we will choose $m_M$, $\sigma_M$ and $k$ as a function of $M$ ($k=k(M)$) that satisfy the following properties:
\begin{align*}
    \lim_{M\rightarrow\infty} \frac{k-m_M}{\sigma_M} \;&=\; s
    \hspace{5mm}\hbox{for some real number $s$}.\\
&\text{and} \\
     \Pr(X_M=k(M)) \;&\sim  \;  \frac{1}{\sqrt{2\pi} \, \sigma_M}  \,\mathrm{e}^{-s^2/2}  
    \hspace{5mm}\hbox{as $M\rightarrow\infty$}
\end{align*}
\noindent 
In our case (theorem \ref{lem8}), we will choose $m_M=\left(1-\frac{1}{R}\right)M$ and $\sigma_M=\sqrt{M/R}$. Fix a real number $s$, we choose the sequence $k(M)$ such as 
 \begin{equation}
  \label{eq:101}
    \lim_{M\rightarrow\infty} \frac{k(M)-m_M}{\sigma_M} =s
\end{equation}
$k(M)$ is provided by the previous Taylor's expansion condition $k(M)= \left(1-\frac{1}{R}\right)M(1+\delta)$ \ \ with \ \ $Min(\frac{1}{R-1},1)>\delta$.\\
\noindent 
The limit (\ref{eq:101}) holds if $\delta = \delta(M)=\frac {s \sqrt{R}}{(R-1)\sqrt{M}}$. In addition, we have the following property 
\begin{align*}
\lim_{M\rightarrow\infty}\delta &=0 & \lim_{M\rightarrow\infty}M{\delta}^3 =\lim_{M\rightarrow\infty}\frac{s^3 R^{\frac{3}{2}}}{(R-1)^3\sqrt{M}} &=0
\end{align*} 
From the function (\ref{eq:100}) and the equivalence (\ref{eq:99}), as $M\rightarrow\infty$, we have:
 \begin{equation*}
 \pi(k) \sim \frac {1} {\sqrt{2\pi} \sigma_M} \mathrm{e}^{\left\{-\frac{1}{2}s^2\right\}} \hspace{10mm}\hbox{$\Box$} \end{equation*} 
 
\section {Conclusion}
\noindent
The dynamics of the infected size of the SIS epidemic model and its distribution at the equilibrium were our main interests in this article. The stability and the equilibrium convergence of the resulting infected size was shown through the sample path simulations. In addition to the dynamics, the stochastic simulations show that when the reproduction number $(R)$ increases for $R>1$, the infected size sample path increases as a whole. These results are not different from the deterministic findings. The distribution of infected size is based on the Birth and Death Markov process. It results from the analysis that the distribution of the infected size is a symmetric-bell shaped curve, the mean and the variance are functions of the reproduction number ($R$). An in-depth analysis of the distribution shows that asymptotic distribution of the infected size follows a Normal distribution with mean $(1-\frac{1}{R}) M$ and variance $\frac{M}{R} $.\\

\section*{acknowledgement}
\noindent 
I would like to express my special thanks to Prof.\ Neal Madras for providing advice and feedback on this article.

 \pagestyle{plain}
\addcontentsline{toc}{chapter}{References}
\bibliographystyle{unsrt}
\bibliography{refer20.bib}

\end{document}